# Small bodies science with the Twinkle space telescope

Billy Edwards
Sean Lindsay
Giorgio Savini
Giovanna Tinetti
Claudio Arena
Neil Bowles
Marcell Tessenyi

# Small bodies science with the Twinkle space telescope

**Billy Edwards,[a,*] Sean Lindsay,[b] Giorgio Savini,[a,c] Giovanna Tinetti,[a,c] Claudio Arena,[a] Neil Bowles,[d] and Marcell Tessenyi[a,c]**

[a]University College London, Department of Physics and Astronomy, London, United Kingdom
[b]University of Tennessee, Department of Physics and Astronomy, Knoxville, Tennessee, United States
[c]Blues Skies Space Ltd., London, United Kingdom
[d]University of Oxford, Atmospheric, Oceanic, and Planetary Physics, Clarendon Laboratory, Oxford, United Kingdom

**Abstract.** Twinkle is an upcoming 0.45-m space-based telescope equipped with a visible and two near-infrared spectrometers covering the spectral range 0.4 to 4.5 $\mu$m with a resolving power $R \sim 250$ ($\lambda < 2.42$ $\mu$m) and $R \sim 60$ ($\lambda > 2.42$ $\mu$m). We explore Twinkle's capabilities for small bodies science and find that, given Twinkle's sensitivity, pointing stability, and spectral range, the mission can observe a large number of small bodies. The sensitivity of Twinkle is calculated and compared to the flux from an object of a given visible magnitude. The number, and brightness, of asteroids and comets that enter Twinkle's field of regard is studied over three time periods of up to a decade. We find that, over a decade, several thousand asteroids enter Twinkle's field of regard with a brightness and nonsidereal rate that will allow Twinkle to characterize them at the instrumentation's native resolution with SNR > 100. Hundreds of comets can also be observed. Therefore, Twinkle offers researchers the opportunity to contribute significantly to the field of Solar System small bodies research. © 2019 Society of Photo-Optical Instrumentation Engineers (SPIE) [DOI: 10.1117/1.JATIS.5.3.034004]

## 1 Introduction

Spacecraft studies of Solar System bodies have increasingly contributed to our knowledge of these objects over recent years. Although *in situ* measurements provide the best means of understanding a target, dedicated lander, orbiting, or fly-by missions are rare and thus remote-sensing missions offer a great chance to observe many objects of interest. Some targets can be viewed by ground-based telescopes at certain wavelengths (e.g., visible) but significant issues are encountered in other bands due to atmospheric absorption, particularly if observing at infrared (IR) or ultraviolet (UV) wavelengths. Additionally, ground-based observations can be affected by weather and atmospheric distortion. Space telescopes avoid these issues and thus are valuable for increasing our knowledge of the universe.

Potential targets for observation within our local stellar environment are diverse and each offers insight into the Solar System as a whole. Asteroids and comets are remnants of the earliest celestial bodies, providing a means of investigating the formation of the planets we know today. Studying the building blocks of the Solar System, as well as the larger bodies that have formed, enhances our understanding of planet formation and evolution. Spectroscopic observations, particularly at visible and IR wavelengths, allow the composition of the surfaces and atmospheres of these objects to be determined and hints of their formation and evolutionary processes to be gleaned.

As these bodies are relatively unaltered since their formation nearly 4.5 Gya, a compositional characterization of small bodies provides a means to identify products of the Solar nebula during the epochs of planetary formation. The implications of such a characterization are far-reaching and, to name a few, include:

- Terrestrial and giant planet formation via understanding the protoplanetary compositions that aggregated into the planets.
- Identification of the source regions for Earth's water and organics.
- How and when differentiation of planetismals, protoplanets, and eventual planets occurred.
- The protoplanetary disk temperature, pressure, and chemical gradient structure.
- The dynamics of the early Solar System via providing observational constraints to early Solar System evolution models such as the grand tack[1,2] and Nice model.[3–8]
- Establishing links between meteorites and asteroids to make inferences on the parent bodies of the meteorites found on Earth.
- The thermal and aqueous alteration history of the small bodies.
- Slow evolutionary processes such as regolith production and space weathering.

Although many small bodies have been discovered, and basic characteristics such as size and orbital parameters are known for many of them, only a small percentage have been characterized through spectroscopy. Many of the spectroscopic studies of asteroids and comets have been with ground-based telescopes such as the Infrared Telescope Facility (IRTF). Additionally, spacecraft has increasingly contributed to our knowledge of these primordial bodies. Fly-by and rendezvous missions (e.g., Rosetta, DAWN, and NEAR Shoemaker) have provided *in situ* observations of a handful of small bodies in extraordinary detail while sample return missions (e.g., Stardust, Hayabusa, Hayabusa 2, and OSIRIS-Rex) can offer unparalleled opportunities for studying the composition of the target bodies. For example, the NIR spectrum of comet 103P/Hartley 2 obtained by the HRI-IR[9] spectrometer during the NASA deep impact extended investigation mission[10] highlighted water-ice absorption and water vapor, organics, and $CO_2$ emission features.[11] Such endeavors allow a few selected objects to be studied in great detail, whereas remote-sensing missions allow for a large number of targets to be studied, albeit in less detail.

Space-based telescopes have been heavily employed for the detection and characterization of small bodies. For example, the Wide Infrared Survey Explorer (WISE) is a medium-class space telescope launched in 2009 that acquired IR images of the entire sky over four bands centered on the wavelengths 3.4, 4.6, 12, and 22 $\mu m$.[12] During the main mission, near-Earth object WISE (NEOWISE), the asteroid hunting section of the project, detected and reported diameters and albedos for >158,000 asteroids, including ∼700 near-Earth objects (NEOs).[13] After the end of the cold mission and several years of hibernation, the spacecraft was reactivated in September 2013.[14] NEOWISE has since detected several hundred more NEOs and thousands of main belt asteroids using the 3.4- and 4.6-$\mu m$ bands, providing albedos and diameters for these newly discovered objects.[15–17]

The Spitzer Space Telescope is, along with Hubble, part of NASA's Great Observatories Program. Launched in 2003, Spitzer carries an infrared array camera (IRAC), an infrared spectrograph (IRS), and a multiband imaging photometer. The IRS was split over four submodules with operational wavelengths of 5.3 to 40 $\mu m$[18] and has not been operational since Spitzer's helium coolant was depleted in 2009. Since the cool phase of Spitzer's mission ended, only the IRAC has remained operational though with reduced capabilities. Thus at the time of writing, no space telescope capable of IR spectroscopy beyond 1.7 $\mu m$ is operational. Spitzer has been used extensively for studying small bodies,[19–22] including the ExploreNEOs program,[23] which was a 500-h survey to determine the albedos and diameters for nearly 600 NEOs during the warm mission phase. The CO and $CO_2$ emission of comets has also been observed with Spitzer.[24] Additionally, the Hubble Space Telescope has provided extraordinary data over a vast range of scientific disciplines including small bodies.[25,26] The Hubble WFC3 camera is currently delivering spectroscopic data at wavelengths shorter than 1.7 $\mu m$.

AKARI was an IR astronomy satellite that was developed by the Japanese Aerospace Exploration Agency and launched in 2006.[27] Over its 6-year life, AKARI surveyed 96% of the sky and contributed to a wide range of IR astronomy, including galaxy evolution, stellar formation and evolution, interstellar media, and Solar System objects. AKARI's infrared camera (IRC) operated from 1.8 to 26.5 $\mu m$[28] and was used for several asteroid surveys. These included a catalogue of albedos and sizes for over 5000 asteroids with measurements in two mid-IR bands (9 and 18 $\mu m$) during the cryogenic phase of the mission.[29] Additionally, a spectroscopic survey of tens of asteroids was conducted over 2.5 to 5 $\mu m$ using the grism of the near-infrared (NIR) channel of the IRC.[30]

Future space telescopes will also offer opportunities for small bodies science and the most anticipated is the James Webb Space Telescope (JWST), which is due to be launched in March 2021. A NIR spectrometer and camera are included within the instrument suite[31] and thus will provide the IR capability that is currently missing (0.6 to 5.3 $\mu m$ and 0.6 to 5.0 $\mu m$, respectively). Additionally, the mid-IR instrument covers the wavelength range 5 to 28 $\mu m$ and is capable of medium resolution spectroscopy.[32] However, a primary issue will be oversubscription and not all the science cases will necessarily need the sensitivity and accuracy of JWST. Hence, while many opportunities exist for Solar System science with this observatory,[33–36] a smaller space telescope would offer an alternative for sources that are too bright to justify the use of JWST.

Upcoming all-sky surveys also offer potential for the characterization of small bodies. These include Euclid, a mission to map the geometry of dark matter in the Universe, which is expected to provide spectra for ∼100,000 asteroids from 0.5 to 2 $\mu m$.[37] Selected earlier this year, the Spectrophotometer for the History of the Universe, Epoch of Reionization and ices Explorer (SPHEREx) is a NASA medium-class explorer mission due for launch in 2023. SPHEREx will observe the whole sky with its 20-cm telescope and is expected to provide spectra of tens of thousands of asteroids over the spectral range 0.75 to 5 $\mu m$ at low resolution ($R \sim 35$ to 140).[38]

Another future space-based telescope that is capable of visible and IR spectroscopy is Ariel, the ESA M4 mission, which will study the atmospheres of ∼1000 transiting exoplanets.[39,40] However, Ariel's mission requirements do not include the ability to track nonsidereal targets and thus its capability for small bodies research may be limited.

Several mission concepts have been studied and submitted to calls by ESA and NASA. Medium class proposals to ESA include CASTAway, which aims to explore the main asteroid belt with a telescopic survey of over 10,000 objects, targeted close encounters of 10 to 20 asteroids, and serendipitous searches to constrain the distribution of smaller objects (<10 m).[41] CASTAway's proposed payload consists of a 50-cm diameter telescope with a spectrometer covering 0.6 to 5 $\mu m$ ($R = 30$ to 100), a thermal imager (6 to 16 $\mu m$) for use during flybys, a visible context imager, and modified star tracker cameras to detect small asteroids. Spectral features related to hydroxyl (OH), water (ice and gas), and hydrated silicates, are either partially or fully obscured in Earth-based observations due to Earth's water-rich atmosphere and the 2.5- to 3-$\mu m$ region is especially prohibitive. CASTAway is designed to be able to detect such features.

Main Belt Comets are objects in stable asteroid-like orbits that lie within the snow line but exhibit comet-like activity[42,43]; 18 such active bodies are currently known. Castalia is a proposed ESA mission to rendezvous with the Main Belt Comet 133P/Elst-Pizarro to perform the first characterization of this intriguing population, making the first *in situ* measurements of the water in the asteroid belt and measuring isotope ratios as well as plasma and dust properties.[44]

Although CASTAway and Castalia were not implemented by ESA, a new small bodies mission, Comet Interceptor,[45] has been accepted for ESA's F Class call for launch with Ariel in 2028. The mission aims to encounter a dynamically new comet (i.e., one that is entering the inner Solar System for the first time) as well as making solar wind measurements. The rendezvous target is likely be a long period comet discovered by the Large Synoptic Survey Telescope (LSST) and characterization of this pristine object will be achieved with a compact, agile set of spacecraft. Although far rarer than long-period comets, Comet Interceptor may also have the capability of encountering an interstellar object passing through our Solar System.

NEOCam is a proposed NASA Discovery class mission and is currently funded for an extend Phase A study. Launching to the Sun-Earth L1 Lagrange point, NEOCam will detect and characterize NEOs with a particular focus on those that could potentially impact Earth (i.e., potential hazardous asteroids). NEOCam's primary science objectives are: (i) assess the present-day risk of NEO impacts, (ii) study the origin and ultimate fate of asteroids, (iii) find suitable NEO targets for future exploration by robots and humans. To facilitate this, NEOCam consists of a 50-cm telescope operating at two photometric channels that are dominated by NEO thermal emission, 4.2 to 5.0 $\mu$m and 6 to 10 $\mu$m, in order to better constrain the objects' temperatures and diameters. NEOCam's field of view (FOV) is significantly larger than that of WISE, allowing the mission to discover tens of thousands of new NEOs with sizes as small as 30 to 50 m in diameter.[46]

Therefore, the small bodies field lacks a space-based remote-sensing mission capable of selectively characterizing thousands of asteroids and comets, through visible and near-infrared (VINR) spectroscopy, in the near-future. Although JWST will undoubtedly provide fantastic insights into a handful of small bodies, without a larger population study, progress in understanding these primordial objects will be slower than hoped.

## 2 Twinkle

The Twinkle Space Mission is a new, fast-track satellite designed for launch in 2022. It has been conceived for providing faster access to spectroscopic data from exoplanet atmospheres and Solar System bodies, but it is also capable of providing spectra of bright brown dwarfs and stars. Twinkle is equipped with a visible (0.4 to 1 $\mu$m) and IR (1.3 to 4.5 $\mu$m) spectrometer (split into two channels at 2.42 $\mu$m). The satellite has been designed to operate in a low Earth, Sun-synchronous orbit.[47,48]

Twinkle is a general observatory managed by Blue Skies Space Ltd. Scientists will be able to purchase telescope time, and Twinkle will provide on-demand observations of a wide variety of targets within wavelength ranges that are currently not accessible using other space telescopes or accessible only to oversubscribed observatories in the short-term future. Although it has been shown that Twinkle has significant capability for characterizing exoplanets,[49] the photometric and spectroscopic accuracy will also be well-suited to observing Solar System objects.

Twinkle is currently entering a phase B design review and thus the technical specifications stated here may change. Twinkle's scientific payload consists of a telescope with a 0.45-m aperture, a fine guidance sensor (FGS), and a VNIR spectrometer, which can be operated simultaneously. The exoplanet light visible spectrometer is a visible spectrometer channel based upon the ultraviolet and visible spectrometer (UVIS) flown on the ExoMars Trace Gas Orbiter. For the Mars application, the UVIS instrument used a dual telescope configuration: nadir (downward viewing of the surface for total atmospheric column measurements) and solar occultation observations (looking at the Sun through the atmosphere from orbit to measure vertical profiles). The telescopes were connected to a single spectrometer via a fiber-optic selector link. This telescope and selector system is not required in the Twinkle application as the spectrometer is positioned in the visible beam of the main Twinkle telescope.

The main modification to the spectrometer design is the use of an alternative grating and associated coatings to optimize the spectral range to the visible to near IR range between 0.4 and 1 $\mu$m with a resolving power of $R \sim 250$.[48] Other planned changes include a minor electronics component change on the detector board and relocation of the main electronics board stack to improve thermal isolation and allow the detector to run at a lower temperature. Changes to the firmware code within the electronics will optimize the operations (e.g., CCD readout modes) and integration times for the Twinkle application.[48] This instrument is referred to as channel 0 (Ch0). For the phase A study, an e2v CCD-230-42 detector was assumed for the visible channel, but this is currently under further discussion.

The design of Twinkle's NIR spectrometer is detailed in Ref. 50. The NIR spectrometer will split the light into two channels (1.3 to 2.42 $\mu$m and 2.42 to 4.5 $\mu$m) to provide broadband coverage while also ensuring appreciable spectral resolution. For shorter wavelengths ($\lambda < 2.42$ $\mu$m), the NIR spectrometer will have a resolving power of 250, whereas for longer wavelengths ($\lambda > 2.42$ $\mu$m) this will be reduced to 60.[50] These are referred to as channels 1 and 2 (Ch1 and Ch2), respectively, and the spectrometer delivers a diffraction-limited image over both channels. In the instrument design, a set of coupling lenslets are adopted to create an image of the aperture on the detectors. These lenses produce several spectra on the detector, with the spectrum from the star slit in the center with three spectra from the background slits on either side.[50] The two channels use different halves of the same detector (assumed to be produced by Selex in the phase A study). Due to this layout, the two IR channels (Ch1 and Ch2) must be read out simultaneously, whereas the visible instrument (Ch0) can be read out independently. This current design features a spectral gap at 1 to 1.3 $\mu$m. A summary of the instrument design in shown in Fig. 1.

The platform has a pointing accuracy of 1 arc min[47] and so an FGS camera is to be used aboard Twinkle to facilitate precise pointing. The current design has a read-out frequency of 1 Hz and the FGS detector has an FOV of $6 \times 6$ arc min. Tip-tilt mirror control electronics will be utilized to keep the target within the slit for the duration of an observation and the pointing precision is expected to be on the order of 100 milliarcseconds (mas). A beam splitter is used to divide light between the visible spectrometer and the FGS, reducing the usable science flux.

The satellite will be placed in a low Earth (600 to 700 km), Sun-synchronous (dawn-dusk) polar orbit with a period of 90 to 100 min. The orientation of the satellite's orbit is constant with respect to the Sun but dictates that Twinkle's instrumentation may have to be retargeted during an orbit to avoid Earth's limb. The boresight of the telescope will be pointed within a cone with a radius of 40 deg, which is centered on the anti-sun vector (i.e., the ecliptic). The field of regard could potentially be expanded to $\pm 60$ deg from the ecliptic for nondemanding targets. Further information on Twinkle, including publications describing the instrumentation, is available on the Twinkle website.[51]

Twinkle's spectrometers cover a similar wavelength range to the proposed CASTAway mission and thus will be capable of detecting many of the same features. This spectral coverage is ideal for the compositional characterization of mafic silicates, hydration features, and organics on asteroid surfaces and within comet comae. A list of the main spectral features of some common minerals is given in Table 1. The sensitivity, pointing stability, and spectral range of the Twinkle observatory are well-suited to study near-Earth and main belt asteroids of all taxonomic types, as well as bright, active comets. Additionally, a space-based observatory offers several opportunities for asteroid characterization that are not possible from the ground, allowing Twinkle to contribute unique spectral data to the small bodies community.

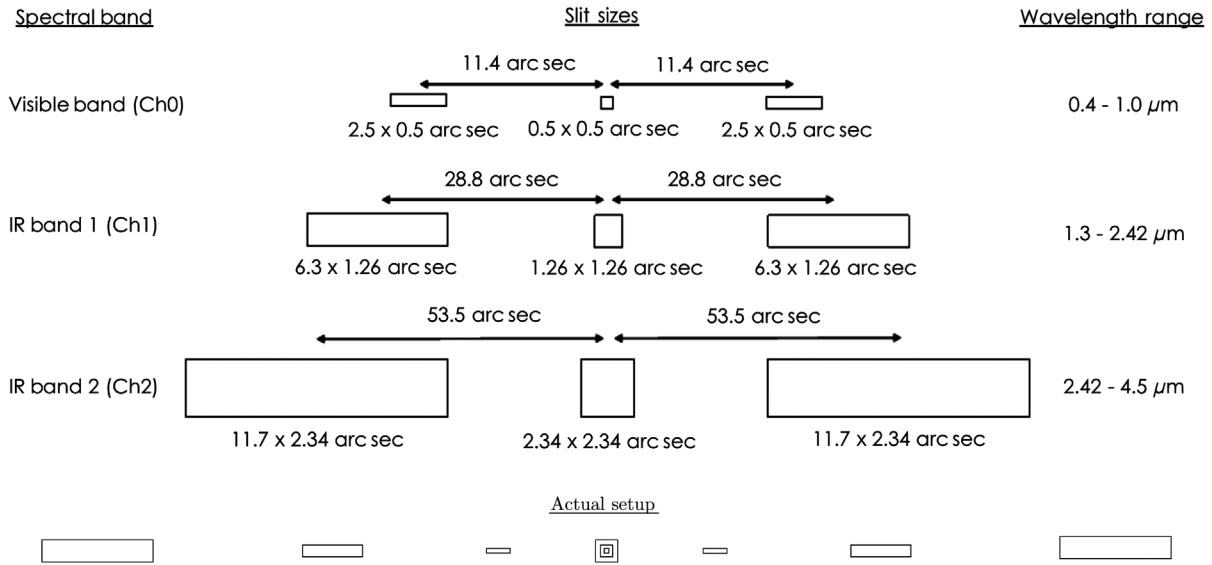

**Fig. 1** Top: angular sizes of Twinkle's star and background slits (to scale) and bottom: the setup and actual separation between them shown to scale. As the star slits are overlapping, all three channels can observe the same target simultaneously.

Hence, while Twinkle observations of asteroids of all taxonomic types would be a great benefit, the ability to obtain spectra outside the water-rich atmosphere of Earth creates a unique opportunity to further our understanding of the primitive asteroids. Currently, a full characterization of these asteroids is strongly limited by contamination of the atmospheric water features, which prevents and complicates the identification of silicate minerals, water ice, OH, and organics contained in the C-complex asteroids.

Here we demonstrate Twinkle's ability to acquire high-fidelity spectral data of many asteroids and comets within the Solar System, provide a description of which spectral features lie within Twinkle's spectral range that make it an ideal small bodies research facility, and discuss which research areas Twinkle could contribute to most significantly.

## 3 Methodology

JPL's Horizons system[52] was accessed for ∼740,000 small bodies defined as NEOs, inner, main, and outer belt asteroids or trojan asteroids. Here NEOs include those that are classified by the Horizons database as Aten, Apollo, and Amor. Atira asteroids are excluded as their orbits are contained entirely within the Earth's orbit and thus are not observable due to the field of regard described in Sec. 2. Mars-crossing asteroids are classified here as inner belt asteroids. Additionally, the physical characteristics of ∼1000 comets were downloaded from the Minor Planet Center.[53] The capability of Twinkle to observe these small bodies has been analyzed using the methods described in Secs. 3.1–3.3. We note that this analysis does not include many comets (over 6000 have been discovered), due to lack of information on key orbital or physical parameters, and that future surveys (e.g., NEOCam and LSST) are expected to find additional small bodies that could be characterized by Twinkle.

### 3.1 Instrument Sensitivity

For many small bodies within the Solar System, such as comets and asteroids, parameters such as albedo, radius, and temperature are not precisely known. Therefore, the analysis described by Edwards et al.,[54] which was used for modeling Twinkle's capabilities to observe major Solar System bodies, cannot be applied. To assess Twinkle's performance when viewing such objects, the flux received has been estimated from the visible magnitude of the body and the methodology for calculating the visible magnitude of small bodies is described in Sec. 3.3. It is assumed that all photons received in Twinkle's visible and IR wavelengths bands are from reflected solar radiation and that a target is small enough to be viewed in its entirety in one observation. The former is valid for Ch0 and Ch1 but provides an underestimation of the flux in the spectral band 2.42 to 4.5 $\mu$m, whereas the latter assumption is true for all but the biggest, brightest objects (e.g., Ceres) that have angular diameters that are greater than the size of Twinkle's slits but are already known to be observable with Twinkle.[54] For each magnitude, the photon flux is calculated per spectral bin, which can then be compared to the sensitivity and saturation limits of Twinkle for a given exposure time. The thermal emission of an asteroid can of course be significant, particularly for NEOs. By ignoring it we are underestimating the number of photons received and thus underestimating the exposure time required. When planning an actual observation with Twinkle, the thermal emission should of course be accounted for to avoid detector saturation. Here, however, we attempt to classify an approximate number of potential targets rather than focusing on any individual objects. To reduce the number of assumptions in the calculation of the flux (e.g., surface temperature, diameter, and albedo), we chose the simplified case of assuming just reflected solar radiation.

The minimum photon flux required to achieve SNR = 100 was calculated for various exposure times to find the sensitivity limit of Twinkle. Additionally, the saturation limit was found by calculating the maximum photon flux that could be observed in each spectral bin. The noise characteristics have been calculated per spectral bin as described in Ref. 54. In line with the phase A design, the detector is assumed to be cooled to 70 K, whereas the telescope has been modeled at 180 K. Excluding the photon noise, the dark current dominates most wavelengths although the telescope noise is pre-eminent at longer wavelengths. With the current design, channel 1 has the lowest noise levels due to the detector dark current.

### 3.2 Pointing and Tracking Restrictions

Twinkle's FGS operates at visible wavelengths and the detailed tracking performance of the FGS will ultimately depend on the platform pointing accuracy. However, it is expected that the wide FOV of the FGS camera will allow bright sidereal targets to be tracked. The ability of Twinkle to track an object varies with brightness and there exists a faintest object that Twinkle can track using the FGS. Current simulations suggest direct tracking will be possible for targets with visible magnitude of 15 or brighter. Further investigation is needed to fully ascertain the capability of the FGS and this will be performed as part of the phase B study. For fainter targets, tracking could be simulated by scanning linear track segments. These linear track segments are linear in equatorial coordinate space; they are commanded as a vector rate in J2000 coordinates, passing through a specified RA and Dec at a specified time. The coordinates of the target can be obtained from services such as Jet Propulsion Laboratory's Horizons System. This method of tracking is by no means simple but has been employed on Spitzer (and will be for JWST). Including such a capability would be nontrivial but, given the current status of the mission, there is time to include and refine this capacity. Here we assume only on-target tracking is used.

The max tracking rate of Twinkle is also subject to further investigation. During the ExploreNEO program, Spitzer targets achieved a max rate of 543 mas/s[23] and JWST will be capable of 30 mas/s.[36] Twinkle's FGS is expected to be capable of tracking Mars (e.g., a rate of 30 mas/s) though its max rate may be higher for brighter targets.

### 3.3 Target Availability

Twinkle has a design life of 7 years but, with no expendables, has the potential to operate for far longer. A precise launch date for the mission is still under discussion: for the purpose of this work a "first light" date of January 1, 2022, was chosen and the following analysis was completed with a mission end date of January 1, 2032. Twinkle's field of regard provides restrictions on the targets that can be observed at a given time. To assess the number of small bodies that enter Twinkle's field of regard, JPL's Horizons system was accessed and ephemeris data obtained for all small bodies over the timescale 2022 to 2032 at one-day intervals. This was compared to Twinkle's field of regard and, when a target could be observed with Twinkle, the visible magnitude $m$ of the body calculated from

$$m = H + 2.5 \log_{10}\left[\frac{d_{S-T}^2 \times d_{O-T}^2}{q(\alpha) \times (1 \text{ AU})^4}\right], \quad (1)$$

where $d_{S-T}$ is the distance between the Sun and the target, $d_{O-T}$ is the distance between the observer and the target, $q(\alpha)$ is the phase integral, and $H$ is the apparent magnitude an object would have if it was at 1 AU from both the observer and the Sun. The apparent magnitude of these bodies was calculated using $d_{S-T}$ and $H$ from the Horizons database. For planetary bodies (with an atmosphere), the phase integral can be estimated as that for a diffuse sphere.[55] However, airless bodies (i.e., asteroids although some have tenuous exospheres such as Ceres[56]) usually reflect light more strongly in the direction of the incident light. This causes their brightness to increase rapidly as the phase angle, the angle between the Sun, the observer, and the target, approaches 0 deg. This opposition effect is dependent upon the physical properties of the body.[57] Therefore, $p(\alpha)$ has been calculated from

$$q(\alpha) = (1 - G)\phi_1(\alpha) + G\phi_2(\alpha), \quad (2)$$

where $G$ is the slope parameter (acquired from Horizons or assumed to be 0.15) and $\phi$ is given by

$$\phi_n(\alpha) = \exp\left[-A_n\left(\tan\frac{\alpha}{2}\right)^{B_n}\right],$$

where $A_1 = 3.332$, $A_2 = 1.862$, $B_1 = 0.631$, and $B_2 = 1.218$.[58,59] This is valid for phase angles below 120 deg. These assumptions, therefore, include phase angle effects but do not account for phase angle-dependent spectral effects. However, except for extreme cases at large phase angles, these effects are relatively minor.

This calculation was performed for three periods (1 year, 3 years, and 10 years), each starting in 2022, to provide estimates of the number of asteroids observable with Twinkle over the mission life but also shorter time spans. By monitoring an asteroid over time, the maximum brightness when observable with Twinkle could be obtained. The visible magnitude was utilized to calculate the number of photons received from the target in each spectral band. The rate of motion at a given time was also calculated to account for the capabilities of the FGS.

The size of the asteroids that could be characterized by Twinkle is of interest and so, if not currently known (i.e., listed in the Horizons database), the diameter (m) has been determined from

$$d = 10^{3.1236 - 0.5 \log_{10}(p_v) - 0.2H}, \quad (3)$$

where $p_v$ is the geometric albedo. For each target, three possible albedo classes are considered that represent a variety of asteroid taxonomies. The albedo classes are defined using average albedos for different taxonomic types: (i) taxonomic types with low-average albedos near 0.05 including the C-complex and P- and D-types (possible X-complex); (ii) taxonomic types with moderate average albedos near 0.20 including the S-complex and K-, L-,(possible X-complex), and M-types (X-complex); and (iii) taxonomic types with high-average albedos near 0.40 including V-type (similar to S-complex) and E-type (X-complex).[60–62]

## 4 Results

### 4.1 Number of Observable Asteroids

By assuming a requirement of SNR = 100 and the discussed instrument characteristics, the capability of Twinkle to observe small bodies is determined by calculating the sensitivity and saturation limits of Twinkle's instrumentation for each spectrometer. These are plotted in Fig. 2 and, if an object lies between these limits for a given exposure time, Twinkle can achieve spectra at the instrumentation's highest resolution with an SNR > 100. At shorter wavelengths ($\lambda < 2.42$ $\mu$m), targets of visible magnitudes brighter than $m_v \sim 13.5$ could be observed at Twinkle's highest spectral resolution in 300 s, whereas for longer wavelengths the magnitude limit for this exposure time is $m_v \sim 12$. As discussed in Sec. 3, thermal emission has been ignored and thus the calculated flux at longer wavelengths is an underestimate for many small bodies.

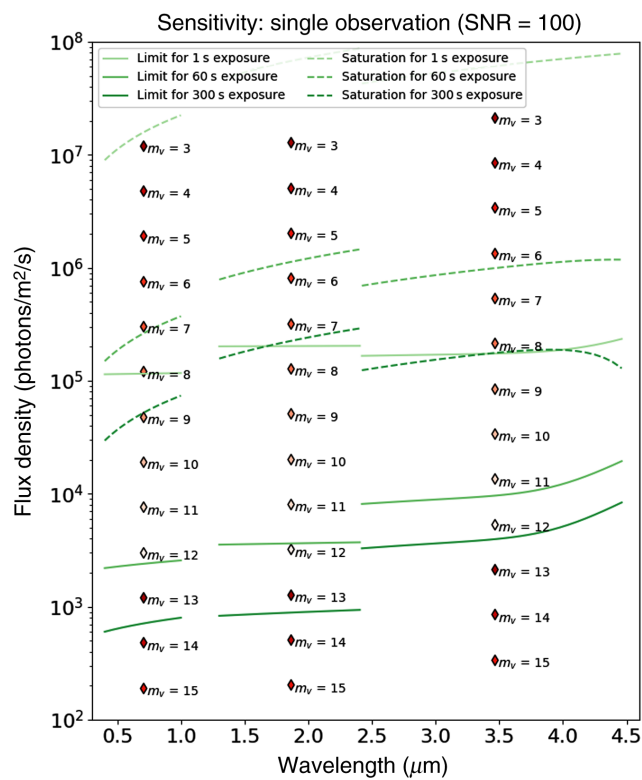

**Fig. 2** For a single observation of a given exposure time, the sensitivity and saturation limits of Twinkle assuming observational parameters of SNR = 100, $R \sim 250$ ($\lambda < 2.42$ $\mu$m), and $R \sim 60$ ($\lambda > 2.42$ $\mu$m). Additionally, the average photon flux received per spectral band at Earth for a small body of a given visible magnitude are plotted.

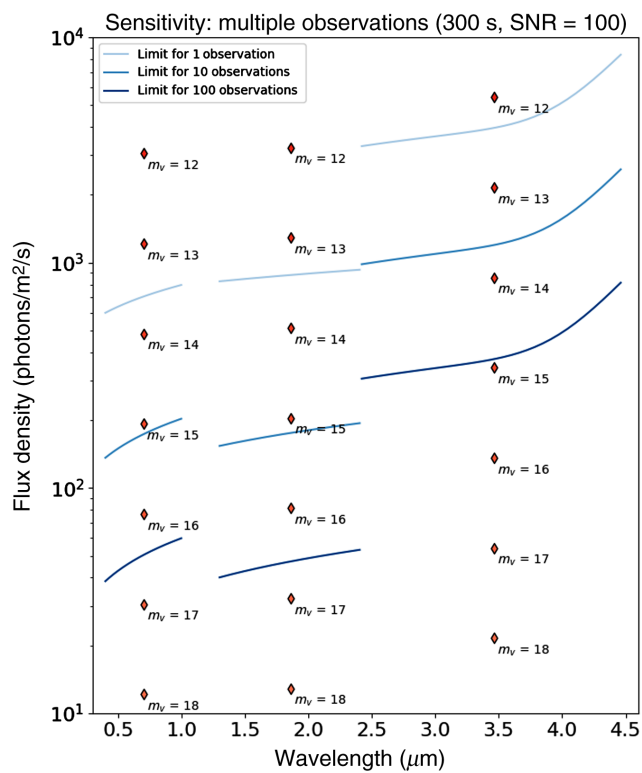

**Fig. 3** For multiple 300-s observations, the sensitivity and saturation limits of Twinkle assuming observational parameters of SNR = 100, $R \sim 250$ ($\lambda < 2.42$ $\mu$m), and $R \sim 60$ ($\lambda > 2.42$ $\mu$m). Additionally, the average photon flux received per spectral band at Earth for a small body of a given visible magnitude are plotted.

By combining multiple observations, the faintest object that could be observed by Twinkle can be improved. We find that by stacking fewer than 100 observations, each with exposure times of 300 s, Twinkle could probe to visible magnitudes of $m_v \sim 15$ to 16.5 (Fig. 3). The sensitivity limit of Twinkle could be further increased by binning down the spectra, reducing the resolution but increasing the number of photons per spectral bin.

The number of asteroids that Twinkle could characterize depends upon the brightness of targets when entering the field of regard, which dictates the possibility of tracking it with the FGS (without the need for linear tracking segments) and the data quality achievable. The cumulative number of asteroids of a given visible magnitude that enters Twinkle's field of regard with nonsidereal rates of <30 mas/s is shown in Fig. 4. We find that several thousands main belt asteroids with a visible magnitude <15 enter Twinkle's field of regard over the time periods considered. Around a hundred outer belt asteroids are bright enough for on target tracking with the current FGS design, as are a handful of Trojans and tens of asteroids in the Inner Belt. Additionally, tens of NEOs could be studied, some of which may be potential Earth impactors, allowing Twinkle to contribute to planetary defence by characterizing bodies in close proximity to the Earth. We note that the $\sim$70 asteroids studied spectroscopically by AKARI are likely to be included within these potential targets for Twinkle.

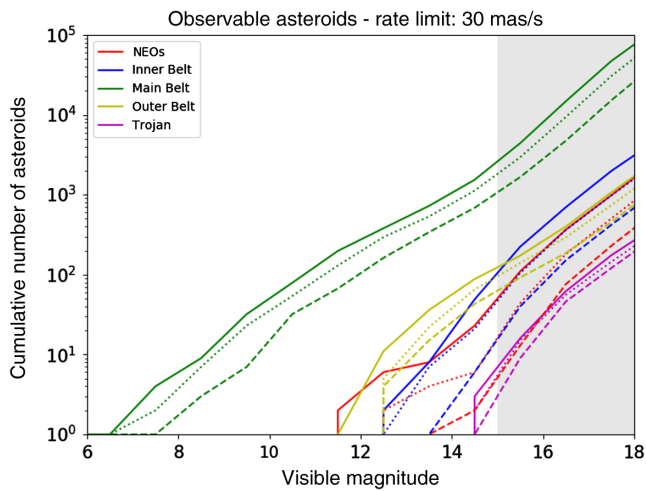

**Fig. 4** Cumulative number of asteroids of a given visible magnitude and type that enter Twinkle's field of regard over several time periods (dashed: 2022 to 2023, dotted: 2022 to 2025, and solid: 2022 to 2032) with nonsidereal rates of <30 mas/s. The gray area indicates the cut-off due to the tracking capability of the current FGS design.

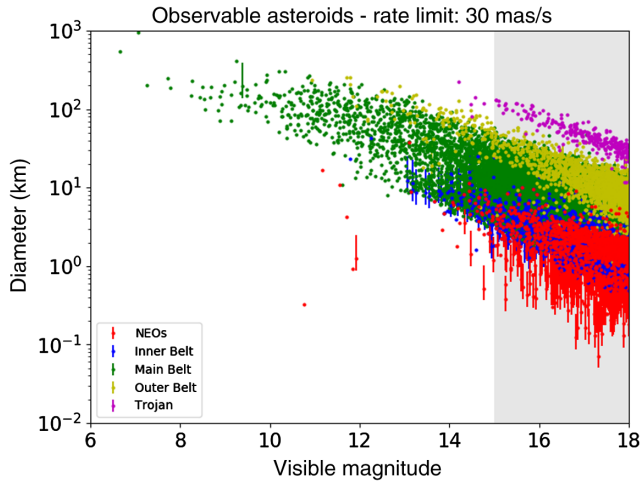

**Fig. 5** The diameters of asteroids that enter Twinkle's field of regard and the max visible magnitude they are observable at with nonsidereal rates of <30 mas/s. The gray area indicates the cut-off due to the tracking capability of the current FGS design.

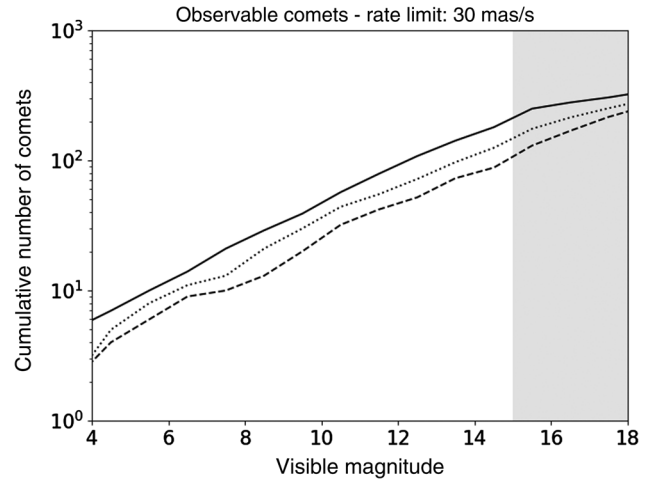

**Fig. 6** Cumulative number of comets of a given visible magnitude that enter Twinkle's field of regard over several time periods (dashed: 2022 to 2023, dotted: 2022 to 2025, solid: 2022 to 2032) with non-sidereal rates of <30 mas/s. The gray area indicates the cut-off due to the tracking capability of the current FGS design.

### 4.2 Size of Potential Observable Asteroids

For each asteroid that enters Twinkle's field of regard over the period 2022 to 2032, the maximum visible magnitude has been calculated as described. Additionally, the diameter, if not already known, has been determined assuming three different albedos (0.05, 0.2, and 0.4). Figure 5 shows the sizes of asteroid that Twinkle could characterize. The plotted value for the diameter is that calculated assuming an albedo of 0.2, whereas the error bars show the change in the diameter if the albedo is between 0.05 and 0.4. We find that the majority of potentially observable asteroids are large (>1 km) but there are also some possible targets with sizes of 100's of metres or less. If tracking via bright stars is employed and data resolution or quality can be sacrificed (i.e., spectral binning of spectra or SNR < 100) then objects fainter than $M_v = 15$ could be observed with Twinkle, which would allow for many asteroids with smaller diameters to be characterized.

### 4.3 Comets

The visible magnitude of ~1000 comets has been monitored over the period 2022 to 2032 and Fig. 6 shows the maximum brightness achieved. Over a decade, ~200 comets are found to be brighter than the current tracking limit of Twinkle's FGS and thus spectra with SNR > 100 could be obtained for these over multiple observations at Twinkle's highest resolution. However, this is likely an underestimate as many known comets have not been included in the analysis, due to lack of a database that provides access to their key orbital and physical parameters, and surveys, such as LSST, which are expected to find many more in the coming years.

## 5 Discussion

With its wavelength coverage, position outside of Earth's atmosphere, instrument performance, and stability, Twinkle is ideally suited to acquire high-fidelity VNIR data for the small bodies of the Solar System. The short exposure times required for many objects mean that Twinkle could observe thousands of small bodies, multiple times, in only a small fraction of the mission life. Although Twinkle is well-suited to investigate asteroids of all types, the most promising contribution is the characterization of primitive asteroids. Twinkle's ability to fully resolve spectral features related to OH, water (ice and gas), and hydrated silicates, which are either partially or fully obscured in Earth-based observations due to Earth's water-rich atmosphere, offers the opportunity to collect the best primitive asteroid VNIR data set to date. Currently, telluric (atmospheric) water features in the spectral data prevents a full characterization of the primitive asteroids and comet comae. Such a characterization is required to determine the composition and mineralogy of these objects and Table 1 (see Sec. 7) contains various minerals with absorption features within Twinkle's spectral range. Without a large database of VNIR spectra free of atmospheric water contamination, our understanding of the composition of the most primitive bodies of the Solar System has been stunted. This has strongly limited the scientific pursuits enumerated in Sec. 1. Here we discuss some of the open questions within the small bodies field and the potential impact Twinkle data could have on these avenues of research.

### 5.1 Characterizing Composition and Mineralogy of Primitive Asteroids

A characterization of the composition and mineralogy of the primitive asteroids (C-complex, some X-complex, D-type, and potentially L-/K-type) remains unconstrained. This is in part due to the fact that the surfaces of these asteroids contain abundant opaques (e.g., amorphous carbon), which lead to the spectra of these asteroid types being, for the most part, featureless. However, the primary problem with characterizing the composition and mineralogy of primitive asteroids using VNIR spectroscopy is that almost all VNIR spectroscopic studies to date have been limited to ground-based observations. The few absorption features observed in primitive asteroids are associated with water and/or hydration (either chemically or physically absorbed), which are contaminated by atmospheric water in ground-based observing campaigns.

### 5.1.1  0.7- and 3-$\mu$m features

Two notable absorption features associated with primitive asteroids are the 0.7- and 3.0-$\mu$m features that are commonly observed in CM chondrites as well as C-complex and M-type asteroids.[63–67] The 0.7-$\mu$m feature is not obscured by atmospheric water, but what compositional information it constrains is still an open question. This feature is often associated with phyllosilicates and attributed to $Fe^{2+} - Fe^{3+}$ intervalence charge transfer.[66,67] Regardless of its association with phyllosilicates, the 0.7-$\mu$m feature is not currently a diagnostic of phyllosilicate mineralogy or composition.

The 3-$\mu$m feature is, in many cases, a complex blend of several different features. The 3-$\mu$m band is associated with hydration and is due to a combination of possibilities: OH, water ice, and water/OH associated with phyllosilicates.[64,68–75] The shape of this feature is dependent on the composition. For example, OH has band positions between 2.7 and 2.8 $\mu$m that vary based on the associated composition (i.e., OH in hydrated minerals).[70] An OH-only 3-$\mu$m feature will have a sharp absorption drop off near 2.7 $\mu$m that transitions to a near-linear return to the continuum level for wavelengths long-ward of the feature minimum. Some studies have shown that the location of the OH band minimum in carbonaceous chondrites is an indicator of phyllosilicates and of the degree of aqueous alteration experienced.[68,71,73,76] A 3-$\mu$m feature due entirely to water ice, on the other hand, will have a broader, more bowl-shaped minimum region.[72,75] The water ice 3-$\mu$m feature is a composite of three absorption bands due to molecular vibrations located near 3.0, 3.1, and 3.2 $\mu$m that shift slightly in wavelength location depending on whether the ice is crystalline or amorphous and as a function of temperature.[72] In addition to the contributions from water and OH, hydrated minerals, such as the phyllosilicates, also have spectral absorption features near 3 $\mu$m with band positions that vary based on mineral species and composition.

The large diversity of 3-$\mu$m band shapes and centers is used to divide the NIR spectra of asteroids with a 3-$\mu$m feature into four spectral groups (the sharp, or "Pallas"-like, rounded or "Themis"-like, "Ceres"-like, and "Europa"-like),[76] and divide the spectra of CM and CI chondrites into three spectral groups, with band positions tenuously associated with degree of alteration and composition of the phyllosilicate serpentine.[75] The relationship between these two groups is still not understood, and there have been no meteorite spectral matches to the Ceres-like, Pallas-like, or Europa-like asteroid spectral groups.[75,76] Efforts to resolve this problem would greatly benefit from spectral data sets of primitive asteroids obtained by a space-based telescope, such as Twinkle, coupled with further laboratory measurements of hydrated minerals and carbonaceous chondrites.

### 5.1.2  3.2- to 3.6-$\mu$m organics feature

The spectrum of C-complex asteroid 24 Themis exhibits a feature spanning 3.2 to 3.6 $\mu$m that has been associated with organic material on the surface.[64,69] This feature is blended with the strong 3-$\mu$m absorption. By fitting a spectral model to the 3-$\mu$m feature that includes water-ice coated pyroxene grains intimately mixed with amorphous carbon, the residual 3.2- to 3.6-$\mu$m feature has been extracted.[64] The shape and position of the residual feature was used to suggest the presence of organic material with $CH_2$ and $CH_3$ functional groups. However, an additional feature centered near 3.3 $\mu$m, indicative of aromatic hydrocarbons, may be required to provide a decent spectral match.

Currently, 24 Themis and 65 Cybele[77] are the only asteroids with spectroscopically confirmed detections of organics. Asteroid 24 Themis is the largest fragment in a dynamical family of over 1600 asteroids located near 3.2 AU from the Sun.[78] There are indications that a significant percentage of them have the 0.7-$\mu$m feature, implying aqueous alteration.[79] Several other studies also suggest that the Themis family has a variety of compositions, based on a large spectral diversity in NIR (1 to 2.5 $\mu$m[80,81]) and mid-IR (5 to 14 $\mu$m[77]) observations. This makes the Themis family a likely candidate to search for organics via an absorption feature near 3.3 to 3.6 $\mu$m. The largest (diameters 50 to 100 km) Themis family members are observable with high SNR for relatively short exposure times, making this group of asteroids targets of interest for Twinkle that could extend the number of detections of organics and our knowledge of organics in the asteroid belt. Additionally, the Themis family contains three of the newly discovered main-belt comets,[43,82,83] where a full 0.4- to 4.5-$\mu$m spectrum uncontaminated by atmospheric water would significantly benefit small bodies science.

### 5.1.3  Additional hydration features

The 1.4- and 1.9-$\mu$m features often used in terrestrial studies, but are yet to be detected in asteroids due to the presence of opaques or contamination by atmospheric water.[70] The 1.4-$\mu$m feature is the first overtone of the OH-band at 2.7 to 2.8 $\mu$m discussed previously. The 1.9-$\mu$m feature is a combination of water ice bending and OH stretching modes. Although these features have yet to be observed in the NIR spectra of asteroids, they are expected to be present in asteroid spectra that exhibit a strong 3.0-$\mu$m feature. Twinkle's position as a space-based telescope offers the opportunity to provide the first detections of these features.

The 2.2- and 2.4-$\mu$m features are OH combination bands that generally appear in pairs,[84] and so they are considered together here. These features are commonly used in Earth and Mars spectral studies to identify phyllosilicates, and they have band center positions that are diagnostic of Al and/or Mg composition.[85] They have been tenuously identified in some CM chondrites and as weak features in a few CI chondrites, but thus far there have only been tentative detections in asteroid spectra.[63,86,87] Twinkle has the potential to identify these compositionally diagnostic features for C- and X-complex asteroids.

## 5.2  Composition of Comet Nuclei and Comae

Comets are considered to be reservoirs of some of the most primitive material in the Solar System. They formed out beyond the $H_2O$ frost-line where ices can condense and become incorporated into growing planetesimals. As such, they contain a plethora of volatile ices, organics, and silicate material that has remained relatively unaltered since the comet forming epoch. This "pristine" quality makes comets an ideal object to study to understand the origin and evolution of our Solar System.

Detecting the solar light reflected by cometary nuclei is a powerful and efficient method for determining their size and for studying their properties. However, this technique requires knowledge of the albedo and the contributions from the comet coma can be difficult to disentangle. Hubble's high spatial resolution has been used to solve this issue.[88,89] However, Twinkle's instrumentation does not provide high spatial resolution and, to characterize the nucleus, will generally have to observe comets near aphelion. This presents an issue as the large separation causes the comet nucleus to be extremely faint. There is also no guaranteed cut-off boundary for cometary activity with many comets know to be active beyond 5 AU[90] and 2P/Encke has been anomalously bright when observed at aphelion.[91]

Hence, Twinkle's capability for determining the size of comet nuclei and detecting water-ice, or other compounds, will require analysis on a case-by-case basis. However, Twinkle will be able to observe the comae of many comets to search for water-ice, water-vapor, $CO_2$, and organics, all of which will add to our understanding of comets and the origins of water and organics in our Solar System.

As a comet nucleus approaches the Sun, the ices near the surface begin to sublimate, liberating material from the surface to form a temporary thin atmosphere of gases and dust (i.e., a comet coma). NIR observations of comet comae frequently reveal the gaseous phase of cometary volatiles: primarily $H_2O$ vapor but also as $CO_2$ and CO gas.[10,92–94] However, over the past decade, there has been a growing number of detections of water-ice (at 2.0 and 3.0 $\mu$m) in the comae of comets made by *in situ* spacecraft or ground-based telescopes.[10,11,92,94–98] Characterizing the composition, size, and structure of these ice grains is a newly emerging field in cometary science. This could be accessible with Twinkle, and it offers the ability to increase our understanding of the initial stages of planet formation, the structure of the early solid grains in the Solar System formation, and the outer disk environmental conditions of the preprotoplanetary disk of gas and dust.

In addition to the water-ice features, Twinkle has the opportunity to detect water vapor (2.7 $\mu$m), organics (3.3 to 3.6 $\mu$m), and $CO_2$ (4.3 $\mu$m) as emission features in the comae of comets.[99–101] Measurements of the 4.3-$\mu$m $CO_2$ feature can be used to derive $CO_2$ abundances in comets. In turn, the abundance of $CO_2$ in comets constrains cometary formation and is the driver of activity on comets, especially at large heliocentric distances that are external to the frost-line. As atmospheric $CO_2$ heavily obscures the 4.3-$\mu$m feature, ground-based studies are unable to observe this feature. A large number of comets were studied as part of the WISE/NEOWISE mission, providing constraints on dust, nucleus size, and the $CO/CO_2$ abundance.[102] Additionally, 23 comets were studied by Spitzer, allowing them to be classified as $CO/CO_2$ "rich" or "poor."[24] However, in both cases, these were broadband measurements making disentangling CO and $CO_2$ emission difficult. AKARI observed 18 comets, constraining the $CO_2$ production, with respect to $H_2O$ production, for 17 of them.[103] Therefore, the total number of comets with observed $CO_2$ features and derived abundances is small. Hence, Twinkle, as a space-based observatory with NIR spectral coverage capable of observing this feature, can potentially provide a highly valuable resource to the cometary science community. Additionally, the LSST is expected to discover ~10,000 comets,[104] some of which will be entering the inner Solar System for the first time. Characterizing these pristine objects, as well as short period comets that have undergone surfaces changes, would allow for a deeper study of comet evolution.

### 5.3 Composition and Mineralogy of S-Complex and V-Type Asteroids

VNIR spectroscopy spanning 0.4 to 2.5 $\mu$m of the stony type asteroids (S-complex and V-type) with mafic mineral compositions primarily of pyroxene and/or olivine are not strongly impeded by atmospheric water, and therefore, there have been numerous spectroscopic studies of these types of bodies. The majority of these studies are performed using the SpeX instrument on the IRTF, a well-subscribed ground-based telescope. Twinkle offers an additional resource to the small bodies community to acquire VNIR spectra of stony asteroid surfaces and provide complementary data. However, Twinkle's current design may limit the characterization of these stony asteroids as discussed in Sec. 5.3.1.

#### 5.3.1 *1.0- to 1.3-$\mu$m spectral gap*

Twinkle's current design has a spectral gap between the visible and IR spectrometers at 1.0 to 1.3 $\mu$m. Twinkle is currently in a phase B design review and thus the instrument characteristics are being reassessed. If a channel that covers the 1.0- to 1.3-$\mu$m region was included, the following types of studies would become possible with Twinkle:

- *Characterization of S-type asteroids to establish links to meteorite analogues*. A major goal of VNIR spectroscopic studies of asteroids is to establish meteorite analogue connections. This requires both high-quality remote-sensing data of asteroids, from a platform such as Twinkle, and several laboratory measurements of meteorites including reflectance spectra and a mineralogical analysis of the meteorites via methods such as electron microprobe or x-ray diffraction. Previous investigations have been successful in establishing such connections.

- *Identification of ordinary chondrite parent bodies*. Another key objective of spectral studies of asteroids is to identify a meteorite analogue, and, in cases where a strong link exists between meteorite type and asteroid type, to leverage that information to identify asteroid families that could represent the parent bodies of those meteorites. Identification of such parent bodies extends our knowledge of the thermal structure of our protoplanetary disk, and of how each of the different OC meteorite groups formed. Hence, using Twinkle to obtain spectra of S-type asteroids in an effort to identify ordinary chondrite parent bodies would add significantly to our understanding of planetary formation, thermal history and evolution of protoplanets, and the dynamical evolution of the Solar System during the protoplanetary disk epochs.

- *Characterization of stony asteroids*. This is limited by the spectral gap. Asteroids with mafic silicates, olivine, and pyroxene on their surfaces have two prominent absorption features near 1 and 2 $\mu$m (Fig. 7) that are diagnostic of silicate mineralogy and composition.[107–112] These two features, often referred to as band I and band II for the 1- and 2-$\mu$m bands, respectively, are commonly used to determine mineralogy (olivine-to-pyroxene ratio) and composition (molar percent of Fe in olivine and pyroxene) of S-complex and V-type asteroids via band parameter analysis studies.[21,105,111–117]

### 5.4 Rotationally Resolved Spectral Data Sets

Based on the exposure time estimates from Fig. 2 and asteroid brightness from Fig. 4, Twinkle will be able to obtain rotationally resolved spectra for a large number of main belt asteroids and some NEOs. Spectral variability is expected for asteroids due to a number of effects, including space weathering, composition, grain size heterogeneity, thermal effects, and viewing aspect (i.e., phase angle of observation). Constraining any of these as the cause of spectral variability offers a tremendous opportunity to further our knowledge on the processes governing the formation and evolution of asteroids. For example, S-type asteroids are susceptible to changes in VNIR spectral slope and band I parameters (depth, center, and area) due to space weathering by irradiation and micrometeorite impacts.[118,119] If subsurface material is brought to the surface via an impact event or rotational fission event (i.e., mass-shedding), this "fresher" material, that has not been processed by space weathering, will have different spectral characteristics than the rest of the regolith on the surface, which would lead to spectral variations as a function of rotation. If detected, this would provide a valuable dataset to understand the rotational evolution and potential disruption of asteroids as well as how space weathering proceeds in different parts of the Solar System.

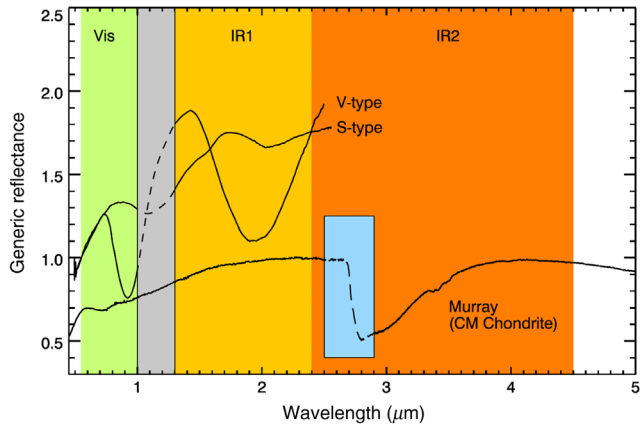

**Fig. 7** Example spectra of potential targets plotted over the visible (Vis) and two near-infrared (IR1 and IR2) channels of Twinkle. The CM chondrite, Murray is representative of primitive, hydrous asteroids. The 0.7- and 3.0-$\mu$m features associated with hydrated primitive asteroids (see Sec. 5.1) are apparent in the spectrum of Murray. The small blue inset window represents the 3-$\mu$m portion of the spectrum obscured by Earth's atmosphere. The Earth's atmosphere can also obscure the spectrum near 1.4 and 1.9 $\mu$m (not shown). The generic S- and V-type asteroid spectra (from Ref. 105) exhibit the 1- and 2-$\mu$m bands used to determine mafic mineralogy and compositions (see Sec. 5.3). The vertical gray bar highlights the spectral gap from 1.0 to 1.3 $\mu$m in the current Twinkle design. The spectrum for Murray was acquired from the RELAB database.[106]

To date, rotational variability in VNIR spectra of asteroids has been observed from *in situ* spacecraft measurements for 951 Gaspra, Ida and Dactyl by Galileo,[120,121] 433 Eros by NEAR,[122] 4 Vesta from the Dawn spacecraft,[123] and ground-based observations,[124] as well as for a handful of NEOs. There have also been suggestions of spectral variability due to surface heterogeneity for other asteroids, but it is likely that these variations are caused by observational effects, such as viewing aspect or poor observing conditions, or different data reduction methods.[105,112] However, as mentioned previously, there are many reasons to expect spectral variability on the surfaces of asteroids. Therefore, it is likely the dearth of confirmed spectral variations due to surface heterogeneity is a result of the reliance on ground-based facilities for VNIR spectroscopy of a large population of asteroids. Considering the relatively short exposure times needed to acquire high signal-to-noise spectra with Twinkle and its position as a space-based observatory, Twinkle could be an ideal telescope to conduct rotationally resolved spectral studies.

## 6 Conclusions

Here we explore Twinkle's capabilities for small bodies science and find that the observatory will have the capability to acquire high SNR spectra for a large variety of asteroid types, including a vast number within the Main Belt. Spectra at Twinkle's highest resolution and with SNR > 100 could be obtained for asteroids brighter than $M_v = 12$ in <300 s. Combining multiple observations, or reducing the observational requirements, will allow many fainter objects to be characterized.

With respect to potential impact, Twinkle's strongest contribution to small bodies science could be the opportunity to investigate the composition of the primitive asteroids that exhibit features associated with hydration (water ice, OH, and phyllosilicates), which to date is an area that has been severely limited due to atmospheric water contamination in ground-based observations. Twinkle also offers the opportunity to study of stony (S-complex and V-type) asteroids. Finally, with respect to asteroid science, Twinkle offers the potential to be the best resource to study rotational variation in spectra of asteroids, which is difficult to do with ground-based telescopes due to Earth's atmosphere generating spectral variations similar to what is expected for asteroids.

Additionally to asteroid science, Twinkle will have the capability of investigating the comae of bright comets, providing valuable data sets on $CO_2$ production and the presence of water-ice and organics in the comae. Therefore, Twinkle potentially provides a resource that would push our understanding of asteroids and comets, and hence the formation and evolution of the Solar System, well beyond its current state.

## 7 Appendix

Various minerals and molecules have absorption features located within the wavelength coverage of Twinkle and some key spectral regions for the characterization of small bodies are highlighted in Table 1.

**Table 1** Main spectral features of some common minerals within the 0.4- to 4.5-$\mu$m spectral region.

| Minerals | Main spectral features ($\mu$m) | Comments |
|---|---|---|
| Carbonates | | |
| Calcite/dolomite | 1.85 to 1.87 | |
| | 1.97 to 2.0 | |
| | 2.12 to 2.16 | |
| | 2.30 to 2.35 | |
| | 2.50 to 2.55 | |
| | 3.40 | |
| | 4.00 | |

Table 1 (*Continued*).

| Minerals | Main spectral features ($\mu$m) | | Comments |
|---|---|---|---|
| Oxides | | | |
| Chromite | | 0.49 | |
| | | 0.59 | |
| | | 1.3 | At the edge of channel 1 |
| | | 2.0 | |
| Spinel | | 0.46 | |
| | | 0.93 | |
| | | 2.80 | |
| Organics | | | |
| e.g., n-alkanes and amino acids | Numerous including: | 1.7 | |
| | | 2.3 | |
| | | 2.4 | |
| Phosphates | | | |
| Apatite | OH—apatite | 1.4 | |
| | | 1.9 | |
| | | 2.8 | |
| | | 3.0 | |
| | F—Cl apatite | 2.80 | |
| | | 3.47 | |
| | | 4.00 | |
| | | 4.20 | |
| Silicates | | | |
| Olivine | | 0.86 to 0.92 | |
| | | 1.05 to 1.07 | Not currently covered |
| | | 1.23 to 1.29 | Not currently covered |
| Pyroxene | Mg—Fe | 0.91 to 0.94 | |
| | | 1.14 to 1.23 | Not currently covered |
| | | 1.80 to 2.07 | |
| | Ca—Mg—Fe | 1.2 | Not currently covered |
| | | 2.0 | |

Table 1 (*Continued*).

| Minerals | Main spectral features ($\mu$m) | | Comments |
|---|---|---|---|
| Feldspar | Na—Ca | 1.1 to 1.29 | Not currently covered |
| Phyllosilicate | e.g., saponite | 1.35 | |
| | | 1.8 | |
| | | 2.3 | |
| | | 2.8 | |
| | e.g., serpentine | 1.4 | |
| | | 1.9 | |
| | | 2.2 | |
| | | 2.9 | |
| | | 0.7 | |
| | | 0.9 | |
| | | 1.1 | Not currently covered |

## Acknowledgments

This work has been funded through the ERC Consolidator grant ExoLights (No. GA 617119) and the STFC Grant Nos. ST/P000282/1, ST/P002153/1, ST/S002634/1, and ST/T001836/1. We have used the JPL Horizons module of the Python astroquery package. Data were also obtained from the JPL Horizons Small Body database and the Minor Planet Center.

**Billy Edwards** previously studied a master's degree in space science and engineering at UCL/MSSL and received his bachelor's degree in physics from the University of Bath. He is a PhD student at the University College London. His research focuses on capability studies of upcoming space-based telescopes for planetary and exoplanetary science. He is leading several studies into the science performance of Twinkle and is also heavily involved in the ESA M4 mission Ariel.

**Sean Lindsay** is a lecturer at the University of Tennessee whose primary research is on determining the mineralogy and relative abundances of dust species for the small bodies of the Solar System. He has developed various tools to reduce and analyze visible, near-infrared, and thermal infrared spectroscopic data for a variety of instruments on ground and space-based observatories.

**Giorgio Savini** has worked at UCL since 2009 on meta-material concepts for satellite optics and mid-/far-infrared modulation techniques for spectroscopy and interferometry. He worked on the testing of optical components and the ground calibration campaign for the Planck satellite's HFI instrument as well as the software pipeline validation for the Spire spectrometer on the Herschel Space Observatory. He is the chief technology officer for Blue Skies Space Ltd. and payload lead for the Twinkle mission.

**Giovanna Tinetti** is a professor of astrophysics at the University College London and the director of the new UCL Center for Space Exoplanet Data at Harwell. She is the principal investigator of Ariel, the European Space Agency's next medium-class (M4) science mission. She is also co-founder and co-director of Blue Skies Space Ltd., which aims at creating new opportunities for science space satellites.

**Claudio Arena** is a PhD student at the University College London and is a part of the Astronomical Instrumentation Group. His research focus is on fine guidance systems (FGS). His research includes simulations of Twinkle's FGS performance, as well as modeling, simulating, and breadboard level testing an innovative FGS design using piezoelectric actuators.

**Neil Bowles** is an associate professor at the University of Oxford. His main research interests are in laboratory measurements that help analyze and interpret data returned from space-based remote sensing and *in situ* instruments for landers. He also works on developing new space-based instrumentation. He is a co-investigator and science team member on numerous ESA and NASA missions including Ariel and Mars Insight.

**Marcell Tessenyi** received his PhD in astrophysics from the University College London in exoplanet spectroscopy. He is the CEO of Blue Skies Space Ltd. and a project manager for the Twinkle mission. He is responsible for the day-to-day programmatic activities of the Twinkle project. His contributions to space instruments include the European Space Agency's M3 candidate mission EChO and the M4 mission Ariel.